\documentclass[final, aps, pra, floatfix, reprint, nofootinbib, citeautoscript, superscriptaddress, balancelastpage, longbibliography, showkeys]{revtex4-2}
\usepackage[T1]{fontenc}
\usepackage[english]{babel}

\usepackage{amsmath}
\usepackage{amssymb}
\usepackage{bm}
\usepackage{wasysym}
\usepackage{textcomp}
\usepackage[normalem]{ulem}

\usepackage{tabularx}

\usepackage[colorlinks, urlcolor=blue, citecolor=blue, linkcolor=blue, pdfstartview=FitH]{hyperref}
\usepackage[all]{hypcap}
\usepackage[dvipsnames]{xcolor}
\usepackage{subfigure}
\usepackage{graphicx}

\def\affiIOFFE{Ioffe\ Institute, 194021 St.~Petersburg, Russia}

\begin{document}

\author{M.\ V.\ Petrenko}
\affiliation{\affiIOFFE}

\author{A.\ K.\ Vershovskii}
\email{antver@mail.ioffe.ru}
\affiliation{\affiIOFFE}

\selectlanguage{English}

\title{Bistability of optical properties of cesium vapor due to collective interaction of alignment and orientation under strong spin exchange conditions}
%\date{\today}
\begin{abstract}
Hydrogen-like alkali atoms with a single valence electron are the most common objects in quantum optics and, at the same time, serve as essential tools of the field. Under conditions of optical pumping, strong spin-exchange and ultra-weak magnetic field (spin-exchange relaxation free mode, SERF), ensembles of such atoms in the gas phase can demonstrate not only the absence of spin-exchange relaxation, but also nonlinear collective effects. We present experimental evidence that the alignment, i.e. the quadrupole momentum, can not only be preserved under SERF conditions, but also coexist and interact with the orientation, i.e. the dipole momentum. We also show that this interaction leads to bistability: a small change in conditions can cause the medium to transition to a different steady state, an effect characterized by hysteresis. The combination of properties of this effect opens up a wide range of applications as optical keys or memory elements with a storage time of hundreds of seconds in tasks of quantum information and cryptography. \end{abstract}

\maketitle
\textcolor{Black}{\section{Introduction}\label{sec:1}}
Progress in quantum optics in recent decades is largely based on the unique properties of alkali atoms, primarily rubidium and cesium. These substances are characterized by spectra  that are  relatively simple, yet complex and diverse enough to implement dozens of schemes for interaction with electromagnetic radiation in radiofrequency, microwave and optical ranges. They can be brought to a nonequilibrium energy state (pumped) by light from available diode lasers. The concentration of these atoms in saturated vapor is easily controlled by changing the temperature of the container (the working cell). Due to these properties, alkali atoms have become the basis of such devices as atomic clocks \cite{Kitching_2018, Knappe_Schwindt_Shah_Hollberg_Kitching_Liew_Moreland_2005}, magnetic field sensors \cite{Budker_Romalis_2007,	 Petrenko_Pazgalev_Vershovskii_2021, Petrenko_Pazgalev_Vershovskii_2023}, rotation sensors \cite{Meyer_Larsen_2014, Vershovskii_Litmanovich_Pazgalev_Peshekhonov_2018}, etc. Most of these devices rely on optical orientation, a first-order (dipolar) moment of angular momentum. A high degree of orientation was achieved by using buffer gases \cite{Franzen_1959}, or anti-relaxation coatings \cite{Bouchiat_Brossel_1966}. In  \cite{Happer_Tam_1977, Appelt_Ben-AmarBaranga_Young_Happer_1999} it was shown that spin-exchange relaxation can be suppressed when most of the colliding atoms are in the same state. An example is the SERF effect, which provides nearly complete suppression of spin-exchange relaxation. This effect is due to the conservation of the total angular momentum of hyperfine sublevels, which under ultra-frequent collisions precess with a common phase \cite{Happer_Tam_1977}. Based on Happer's theory, zero magnetic field SERF (spin-exchange relaxation free) sensors were created \cite{Kominis_Kornack_Allred_Romalis_2003, Ledbetter_Savukov_Acosta_Budker_Romalis_2008}, followed by non-zero field sensors with partial suppression of spin-exchange relaxation \cite{Petrenko_Pazgalev_Vershovskii_2021,
Scholtes_Schultze_IJsselsteijn_Woetzel_Meyer_2011, Schultze_Schillig_IJsselsteijn_Scholtes_Woetzel_Stolz_2017}.

\begin{figure*}[!t]  
\includegraphics[width=\linewidth]{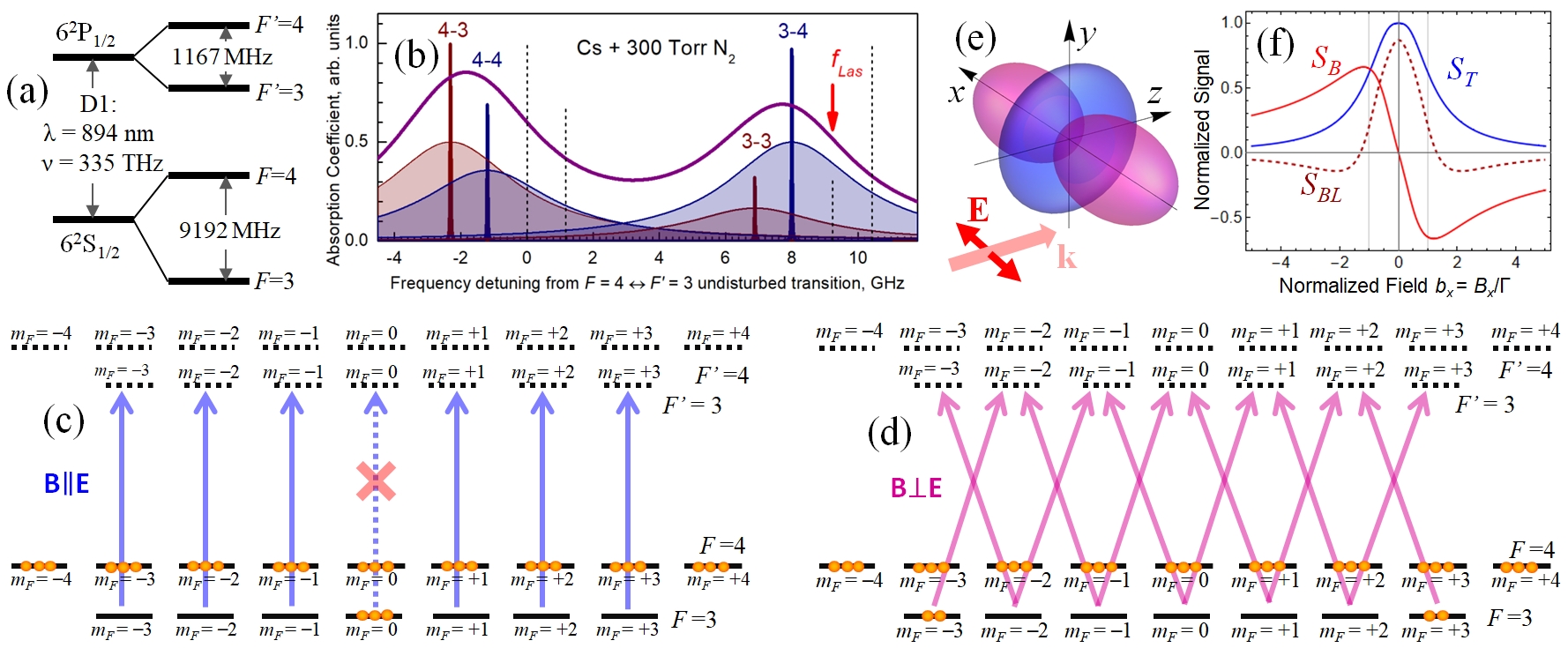}
	\caption{\textcolor{Black}{(a) Energy diagram of the lower levels of Cs; (b) calculated absorption coefficient in a thin layer on the $D_1$ line of Cs in the presence of 300 Torr N$_2$: the profiles corresponding to individual lines are filled in color, the thick purple line is the overall absorption profile. Solid vertical lines are the centers of the profiles shifted by nitrogen pressure, dashed vertical lines are the centers of the undisturbed profiles. The red arrow indicates the laser radiation frequency in our experiment; (c),(d) schemes of pumping with linearly polarized light resonant to the $F = 3 \rightarrow F' = 3$ transition: (c) the external field $\bf{B}$ is parallel to the electric vector $\bf{E}$ of the light wave, the alignment is negative, (d) the field $\bf{B}$ is perpendicular to $\bf{E}$, the alignment is positive; (e) spatial distribution of moments during pumping with linearly polarized light: negative alignment ("donut") and positive ("dumbbell"); (f) calculation of alignment signals in the presence of a small $B_y$ field according to Eqs.\eqref{eq:eq1},\eqref{eq:eq2}: $S_T$ is the absorption signal, $S_B$ is the polarization rotation signal, $S_{BL}$ is the result of lock-in amplification of the modulated $S_B$ signal.}}\label{fig1}
\end{figure*}
Much less attention has been paid to the second-order (quadrupole) moment: the alignment  characterized by a zero value of the average angular momentum. It is known, however, that quantum magnetometers based on alignment can be characterized by significant advantages: the absence of dead zones \cite{Ben-Kish_Romalis_2010, Wang_Wu_Xiao_Wang_Peng_Guo_2021}, small orientation error \cite{Hovde_Patton_Versolato_Corsini_Rochester_Budker_2011, Zhang_Kanta_Wickenbrock_Guo_Budker_2023}, and low drift \cite{Rosner_Beck_Fierlinger_Filter_Klau_Kuchler_Rosner_Sturm_Wurm_Sun_2022}. The description of the alignment effects in alkali metals is given in \cite{Akbar_Kozbial_Elson_Meraki_Kolodynski_Jensen_2024, Weis_Bison_Pazgalev_2006}, and developed for quasi-stationary cases in \cite{Meraki_Elson_Ho_Akbar_Kozbial_Kolodynski_Jensen_2023}. The theoretical definition of higher-order moments is given in \cite{Blum_2012, Omont_1977}. In \cite{Breschi_Weis_2012, LeGal_Lieb_Beato_Jager_Gilles_Palacios-Laloy_2019, Meraki_Elson_Ho_Akbar_Kozbial_Kolodynski_Jensen_2023}, zero-field, alignment-based magnetometers that do not use the SERF effect were proposed. In our recent work \cite{Petrenko_Vershovskii_2025}, however, it was shown that the SERF mode can also be achieved in alignment, although a theoretical explanation for this effect has not yet been given. The elements of the theory developed in \cite{Breschi_Weis_2012, Happer_Tam_1977, Meraki_Elson_Ho_Akbar_Kozbial_Kolodynski_Jensen_2023} that are most relevant for the present study are also summarized in \cite{Petrenko_Vershovskii_2025}.
The present work is a continuation of \cite{Petrenko_Vershovskii_2025}. It is focused on anomalous phenomena arising at high concentrations of atoms in zero magnetic fields, when a small part of circular polarization, corresponding to an ellipticity angle $\chi \le 1^\circ$, is mixed with the linear polarization of the pump light. Bistability in atomic vapors under optical pumping was first investigated in \cite{Fortson_Heckel_1987}, then in \cite{Klipstein_Lamoreaux_Fortson_1996, Andalkar_Warrington_Romalis_Lamoreaux_Heckel_Fortson_2002, Chen_Xiang_Jin_Xiao_Peng_Guo_2024}. It was shown that under the influence of spin-exchange, collective effects arise, as a result of which the atoms tend to be in one "\textcolor{Black}{stretched}" state \cite{Happer_Tam_1977, Appelt_Ben-AmarBaranga_Young_Happer_1999} - which is also the essence of the SERF effect. By "\textcolor{Black}{stretched}" we mean a state in which spin-exchange relaxation is significantly suppressed due to the predominance of atoms in the same spin state; in the case of alignment, such a state can be, for example, a superposition of states with opposite angular momentum projections. The present work differs fundamentally from \cite{Fortson_Heckel_1987, Klipstein_Lamoreaux_Fortson_1996, Andalkar_Warrington_Romalis_Lamoreaux_Heckel_Fortson_2002, Chen_Xiang_Jin_Xiao_Peng_Guo_2024} in that 
it demonstrates the possibility of the coexistence of two \textcolor{Black}{stretched} ensembles, as well as switching the state of one ensemble (alignment) by influencing it with a second ensemble (orientation).

\textcolor{Black}{The remainder of the paper is organized as follows: Section \ref{sec:2} provides a brief background, Section \ref{sec:3} describes the experimental setup, Section \ref{sec:4} presents the experimental results, and Section \ref{sec:5} discusses them.}

\textcolor{Black}{\section{Basic principles}\label{sec:2}}
\textcolor{Black}{The primary object of study in this work is cesium atoms in the gas phase, excited by the resonant light of the $D_1$ line (Fig.~\ref{fig1}(a)). The atoms are surrounded by buffer gas (nitrogen) molecules, which perform several functions: they quench the fluorescence of excited Cs atoms, ensure complete mixing of the Zeeman sublevels of the excited state, and shift and broaden the spectral absorption lines of Cs. At a buffer gas pressure of 200 Torr or more,  the spectral profiles corresponding to transitions to two hyperfine levels of the excited state are non-resolved, and partial overlap of the ground state profiles is observed (Fig.~\ref{fig1}(b)).}

\textcolor{Black}{Phenomena arising when alkali atoms are exposed to circularly polarized resonant light are widely known: photons transfer angular momentum to the atoms, the uniform (under normal conditions) population distribution of Zeeman sublevels is disrupted, and the medium becomes oriented. Under strong pumping, most of the atoms can gather at a single level with the maximum angular momentum projection modulus; this state is referred to as "stretched". At high atomic concentrations, spin-exchange processes begin to dominate, and in zero magnetic fields (we will use this term for fields in which Larmor precession occurs more slowly than relaxation), this dominance leads to the atoms combining into a single ensemble characterized by the maximum angular momentum. Due to the law of conservation of angular momentum in this state, spin-exchange processes no longer destroy the spin state, and the rate of spin-exchange relaxation ultimately drops to zero (SERF mode).}

\begin{figure*}[!t]  
\includegraphics[width=\linewidth]{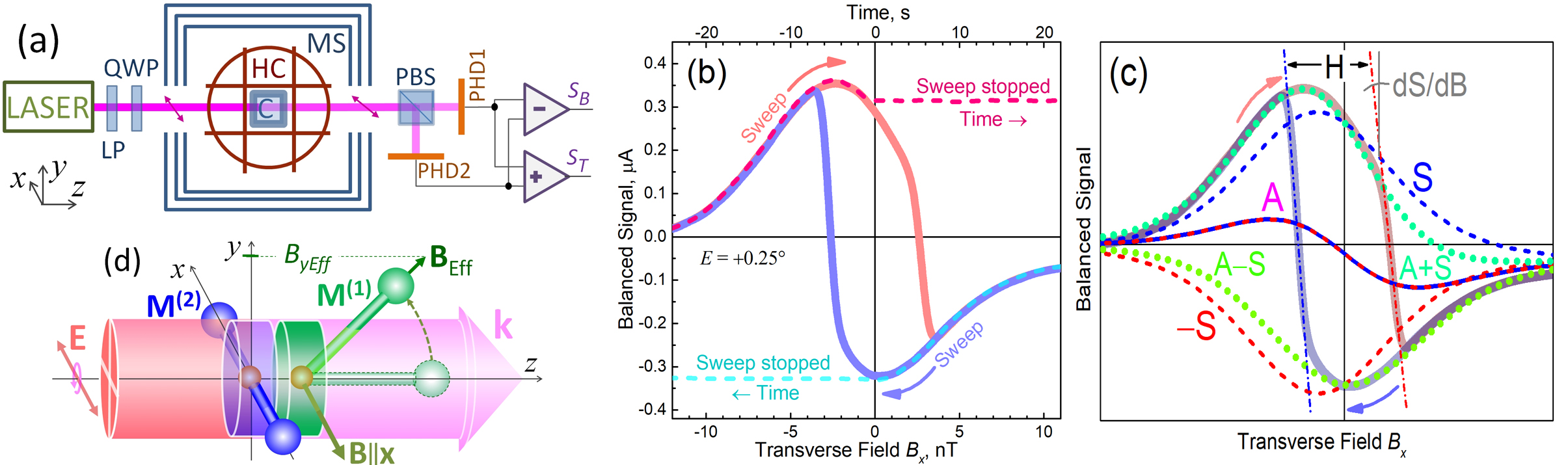}
	\caption{(a) Block diagram of the setup: MS - magnetic shield, LP - linear polarizer (Glan prism), QWP - quarter-wave plate, HC - Helmholtz coils system, C - cell, PBS - polarization cube, PHD1, PHD2 - photodiodes; (b) example of an experimental recording demonstrating the effects of bistability (dashed lines) and hysteresis (solid lines); a digital low-pass filter is applied to the data to eliminate high-frequency interference. (c) model used to explain the results. $\text{A}$ - antisymmetric contour, $\text{S}$ - symmetric contour, $\text{H}$ - hysteresis, $\text{A-S}$ and $\text{A+S}$ - resulting envelopes. Solid lines - experimental data (the same recording as in (b); (d) schematic representation of the moments $\bf{M}^{(1)}$ (orientation) and $\bf{M}^{(2)}$ (alignment) created by weakly elliptical spatially non-uniform optical pumping, and the dynamics of  $\bf{M}^{(1)}$ in the presence of an external field $\bf{B}||\bf{x}$.}\label{fig2}
\end{figure*}

\textcolor{Black}{Far less attention has been given to the alignment effects that arise when alkali atoms are exposed to linearly polarized light. Alignment is defined as the quadrupole component of the angular momentum, characterized by a symmetric distribution over Zeeman sublevels with zero mean value.  Examples of schemes  of pumping on  $F = 3 \rightarrow F' = 3$  transition of the $D_1$ line for two directions of the field are shown in   Fig.~\ref{fig1}(c),(d). It should be noted, however, that due to the gas broadening, pumping  at high gas pressures always occurs at two, or even four, hyperfine transitions. The final result is determined by the ratio of the corresponding pump rates.}

\textcolor{Black}{In \cite{Meraki_Elson_Ho_Akbar_Kozbial_Kolodynski_Jensen_2023}, a theory describing the alignment signals in zero magnetic field is offered, and it is shown that the transmission signals $S_T$ and polarization rotation signals $S_B$ in linearly polarized light transmitted through an atomic medium are described by the expressions}

\begin{equation}
 \textcolor{Black}{S_{T} \propto \frac{1+ \left(b_{yz}^{2}-2b_{x}^{2} \right)^{2} + 2b_{yz}^{2} + 5b_{x}^{2}}{\left(1+ b_{x}^{2} + b_{yz}^{2} \right)\left(1+ 4b_{x}^{2} + 4b_{yz}^{2} \right)}},
    \label{eq:eq1}   
\end{equation}
   
\begin{equation}
   S_{B} \propto \frac{b_{z} \left(1+4b_{x}^{2} +b_{yz}^{2} \right)-b_{x} b_{y} \left(1+\textcolor{Black}{4b_{x}^{2}} -2b_{yz}^{2}\right)}{\left(1+b_{x}^{2} +b_{yz}^{2} \right)\left(1+4b_{x}^{2} +4b_{yz}^{2}\right)},
    \label{eq:eq2}
  \end{equation}
  
\noindent where $b_i = \gamma B_i/\Gamma$, $i = x, y, z$, $b_{yz}^2 = b_{y}^2+b_{z}^2$, and $\gamma$ is the gyromagnetic ratio; \textcolor{Black}{$\Gamma$ is the total relaxation rate due to the following factors: collisions with the buffer gas, with the cell walls, with other Cs atoms (spin-destructive collisions), as well as spin-exchange collisions under conditions of incomplete suppression of spin-exchange relaxation in a nonzero magnetic field, and absorption of pumping light. As follows from Eqs.\eqref{eq:eq1},\eqref{eq:eq2} (and illustrated in Fig.~\ref{fig1}(f)), the observed resonance width is close to the relaxation rate $\Gamma$, but, strictly speaking, is not equal to it.}

\textcolor{Black}{Eqs.\eqref{eq:eq1},\eqref{eq:eq2} are derived under the assumption that the symmetry of the system is determined by the direction of the electric component of the light wave that causes the alignment of cesium atoms (in our case, the $x$-axis, Fig.~\ref{fig1}(e)). Accordingly, this direction is chosen as the quantization axis. However, when scanning the magnetic field, introducing sufficiently strong bias fields, or increasing the beam ellipticity, the symmetry of the system can (and should) change, requiring a change in the basis.}

\textcolor{Black}{To improve the signal-to-noise ratio in experiments, a lock-in amplification procedure is typically used: one of the parameters (e.g., the $B_x$ field) is modulated, and the recorded signal is demodulated by multiplying it by the modulation signal. If both the modulation frequency and amplitude are small compared to the magnetic resonance linewidth, this procedure is equivalent to taking the derivative with respect to the $B_x$ field. As a result of this procedure, the antisymmetric $S_B$ signal contour is transformed into a symmetric one (Fig.~\ref{fig1}(f)). From now on, when discussing polarization rotation signals, we will refer to the result of lock-in amplification.}

In the presence of a small (not exceeding the resonance width) field $B_y$, the demodulated $S_B$ signals are characterized by a symmetrical shape close to Lorentz (Fig.~\ref{fig1}(f)); their sign is determined by the sign of $B_y$. As the first approximation, these signals can be fitted by a function of the form $y_0 +1/(1+x^2)^2$; the degree at the denominator is additionally doubled due to \textcolor{Black}{lock-in amplification}.
In the presence of the $B_z$, these signals are characterized by an antisymmetric form close to the dispersion form; the sign of the signals is determined by the sign of $B_z$. When ellipticity is introduced, the alignment is partially transformed into orientation. The \textcolor{Black}{demodulated} orientation signals are characterized by an antisymmetric form; the sign of the signals depends only on the sign of ellipticity. The contour of the total signal is the sum of the symmetric and antisymmetric contours.

\begin{figure*}[!t]  
	\includegraphics[width=\linewidth]{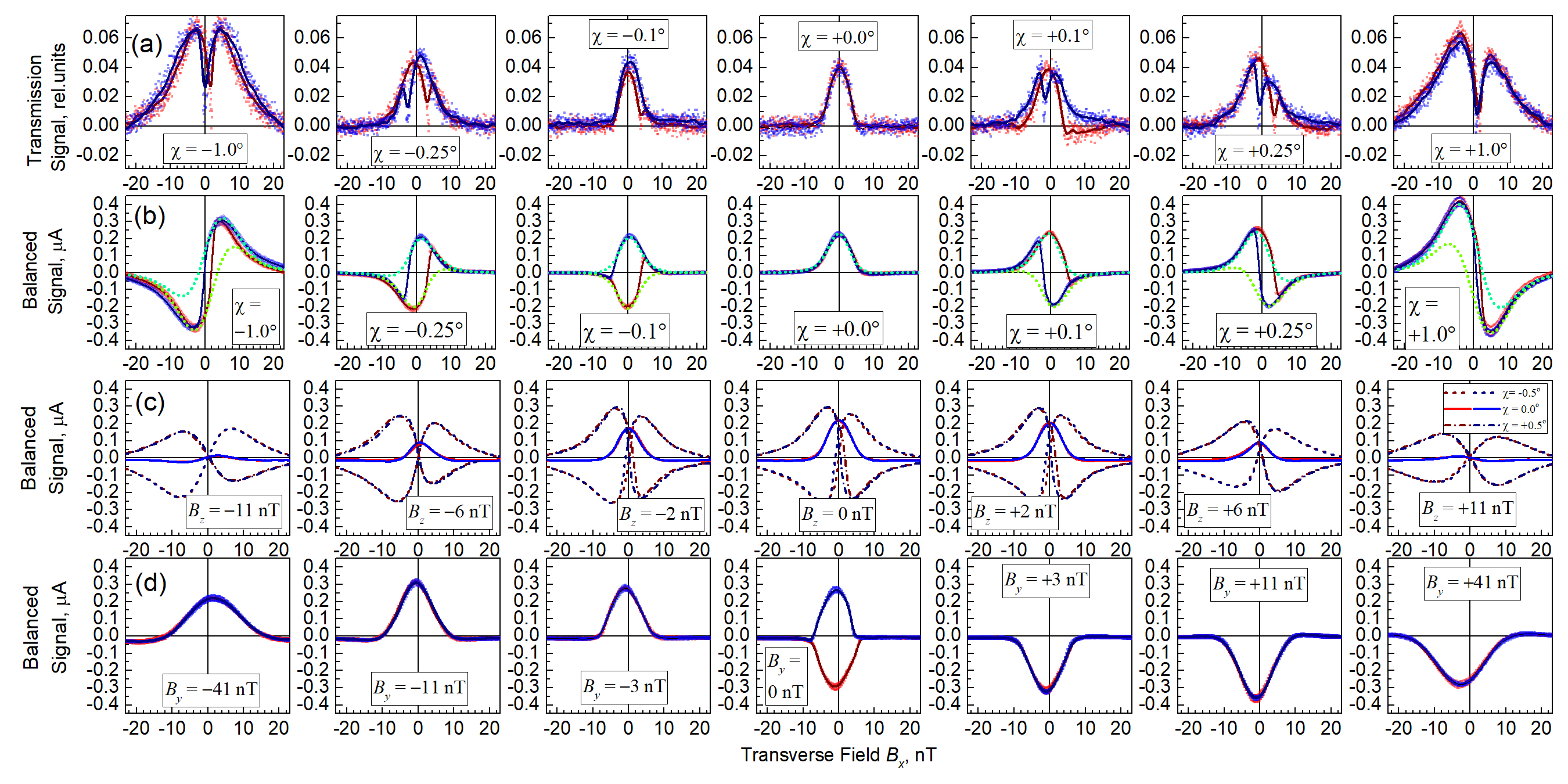}
	\caption{Examples of recordings (a) of absorption signals; (b-d) of \textcolor{Black}{demodulated} polarization rotation signals. In fragments (a-b), the pump ellipticity is varied, (c) the $B_z$ field component is varied at three ellipticity values, (d) the $B_y$ field component is varied at an ellipticity close to zero. In fragment (b), the simulation results are shown as large dots. Fragments (a, b, d) show the original recordings (dots) and the result of digital low-pass filtering (lines). In fragment (c), only the result of digital filtering is shown to avoid overloading the figure.}\label{fig3}
\end{figure*}

These conditions are fully met in our experiment with one addition: in the presence of ellipticity in the pumping light during the scanning of the $B_x$ field, new features arise in the $S_T$ and $S_B$ signals - namely, rapid changes in the signal level, characterized by hysteresis. \textcolor{Black}{In the $S_B$ signal, these rapid changes appear as switching between two stable levels.} In the absorption signal $S_T$ only transient processes that look like extremely narrow (1-2~nT) dips in  background signal are visible (see Fig.~\ref{fig3}(a),(b)). The hysteresis value decreases with increasing ellipticity, as well as with the appearance of additional magnetic fields.

\textcolor{Black}{\section{Experiment}\label{sec:3}}
The block diagram of the experimental setup previously described in \cite{Petrenko_Vershovskii_2025, Petrenko_Pazgalev_Vershovskii_2024} is shown in Fig.~\ref{fig2}(a). A cell containing, in addition to Cs, $P_{N2} \approx$ 300~Torr of nitrogen was chosen for the study. \textcolor{Black}{The cell has a cubic shape and an inner size of $5\times5\times5$~mm$^3$, with several mg of liquid cesium located in a sprout at the bottom. The cell is placed in an oven, which is located in a three-coordinate system of Helmholtz coils in a magnetic shield. The temperature of the sprout is measured with a semiconductor thermistor and then corrected based on preliminary measurements: in a relatively strong ($\sim$10~$\mu$T) magnetic field, the dependence of the magnetic resonance width in the reference cell on temperature was measured, and a correction for the temperature gradient in the thermostat was calculated.}

The pump-detection light propagates along the $z$ axis, the electric component $\bf{E}$ of the pump is directed along the $x$ axis. \textcolor{Black}{The ellipticity measurements are performed using a Thorlabs PAX5710IR1-T-TXP polarimeter.} The VitaWave external cavity diode laser  is tuned to the center of the $F = 3 \rightarrow F' = 3$ $D_1$ transition of the Cs line (895~nm) in a reference vacuum cesium cell. The signals are detected by a balanced photodetector, which allows simultaneous recording of both the transmission signals $S_T$ (by summing the photodiode currents) and the polarization rotation signals $S_B$ (by subtracting the currents). Thus, simultaneous detection of the absorption and polarization rotation of the pump light is performed. The rotation of polarization in the case of orientation is due to circular birefringence \cite{Budker_Gawlik_Kimball_Rochester_Yashchuk_Weis_2002}, and in the case of alignment - to linear dichroism \cite{Meraki_Elson_Ho_Akbar_Kozbial_Kolodynski_Jensen_2023, Petrenko_Pazgalev_Vershovskii_2025, Petrenko_Pazgalev_Vershovskii_2024}. In this case, in zero fields $S_T \sim  \rho_0^{(2)}$ and $S_B \sim \rho_{\pm1}^{(2)}$, where $\rho_0^{(2)}$ and $\rho_{\pm1}^{(2)}$ are the zero and first alignment multipoles, respectively \cite{Breschi_Weis_2012,
Meraki_Elson_Ho_Akbar_Kozbial_Kolodynski_Jensen_2023}. 

All the results presented below were obtained in the cell temperature range $T_c = 145-150 ^\circ $C, which corresponds to an atomic concentration of $1.8-2.2 \times 10^{14}$~cm$^{-3}$ \cite{Margrave_1964}. The pump light power was $P_{\text{in}} \approx $ 2~mW with a beam area of 0.1~cm$^2$.   \textcolor{Black}{The optical thickness of the cell for linearly polarized light was about 5.5, while for circularly polarized light it was about 2.3. This means that linearly polarized light bleaches the atomic medium only to a small extent, while circularly polarized light allows for high degrees of brightening of the medium to be achieved, since it creates a stretched state over a significant area of the cell. The data of our experiment, however, were obtained in the range of small ellipticity values, at which the degrees of influence on the medium of the two components of light -- linearly and circularly polarized -- were comparable. More details are given in Appendix~\hyperref[appendix:A]{A}.}

\textcolor{Black}{The hierarchy of relaxation and exchange processes under these experimental conditions is as follows: spin-exchange interactions predominate, with a rate of approximately 1.3$\times$10$^5$~s$^{-1}$, or 21~kHz. The average optical pumping rate over the illuminated region of the cell is approximately 900~s$^{-1}$, corresponding to an average light-induced broadening of 140~Hz. However, the signals recorded in SERF mode are primarily due to atoms in states that do not interact with light, and such atoms predominate in the SERF spatial regions. This means that in these regions, light is absorbed only by atoms removed from the SERF state by diffusion or by destructive collisions, namely, collisions with the buffer gas, the walls, or other Cs atoms (spin-destructive collisions) -- or as a result of magnetic field modulation, which allows these signals to be observed. The combined rate of all these processes in our experiment is $10-20$~Hz. Thus, in the SERF region, light-induced broadening does not exceed this value. The sum of the above dark relaxation rates with the pumping rate determines the total resonance zero-field width, which in our experiment is $8-10$~nT HWHM. We note once again that such widths are observed, including in typical alignment signals under the linearly polarized pumping, which confirms the conclusion \cite{Petrenko_Vershovskii_2025} about the possibility of the SERF mode in alignment.}

All the records presented below were obtained by scanning the magnetic field component $B_x$ at different values of the components $B_y$, $B_z$ and the pump ellipticity angle $\chi$. The $B_x$ field was modulated by a sine wave with an amplitude of 2.5~nT and a frequency of 5 Hz; the $S_B$ signals were obtained as a result of lock-in amplification. Some records were duplicated without modulation, which showed that modulation does not significantly affect the signal shape \textcolor{Black}{(save for the fact that the lock-in amplification transforms symmetrical contours into asymmetrical ones, and vice versa).} A typical record of the $S_B$ signal obtained at the pump ellipticity angle $\chi = 0.25^\circ$ is shown in Fig.~\ref{fig2}(b). The recordings in which scanning was stopped upon reaching the threshold value $B_x= 0$  are likewise shown there; they confirm that the state of the system is indeed bistable within tens (and even hundreds) of seconds. In fact, the system in our experiment remains in a stable state up to 300~s -- as long as the field drift in the shield does not exceed $1-3$~nT.

\textcolor{Black}{\section{Results}\label{sec:4}}
Analysis of the obtained records (Fig.~\ref{fig3}) allows us to conclude that these features are due to the fact that under certain conditions the alignment signal is inverted.  This conclusion is confirmed by the results of modeling, in which the envelopes of these signals are represented by the sum and difference of the antisymmetric and symmetric contours (Fig.~\ref{fig3}(b)). The model is illustrated in Fig.~\ref{fig2}(c) using the the record (Fig.~\ref{fig2}(b)) as an example.

\begin{figure}[!t]  
	\includegraphics[width=\linewidth]{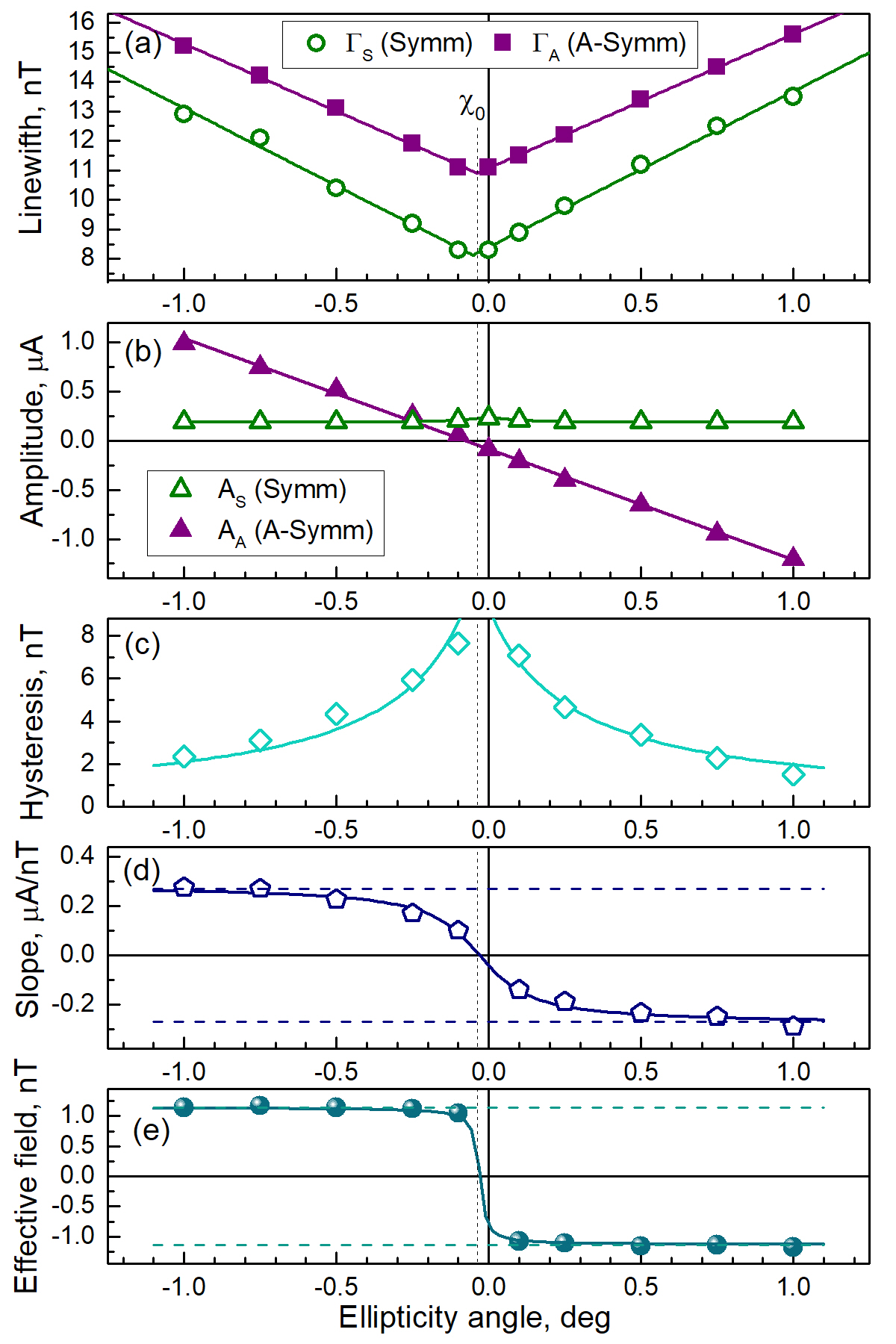}
	\caption{Results of fitting the data in Fig.~\ref{fig3}(b) to the "antisymmetric $\pm$ symmetric contour" model: (a) widths of both contours. Solid lines are the linear approximation; (b) amplitudes of both contours. Solid lines are the linear approximation and Lorentzian contour approximation, (c) hysteresis in magnetic field units. Solid line is the hyperbolic approximation; (d) \textcolor{Black}{maximum values of the slope $dS_B/dB_x$ of the transient process, that is, the change in the $dS_B$ signal during scanning $dB_x$}. Solid line is the arctangent approximation; (e) value of the effective field that could lead to a reversal of the alignment moment by $\pi$ during the time corresponding to the duration of the transient process. Solid line is the arctangent approximation.}\label{fig4}
\end{figure} 
\begin{figure}[!t]  
	\includegraphics[width=\linewidth]{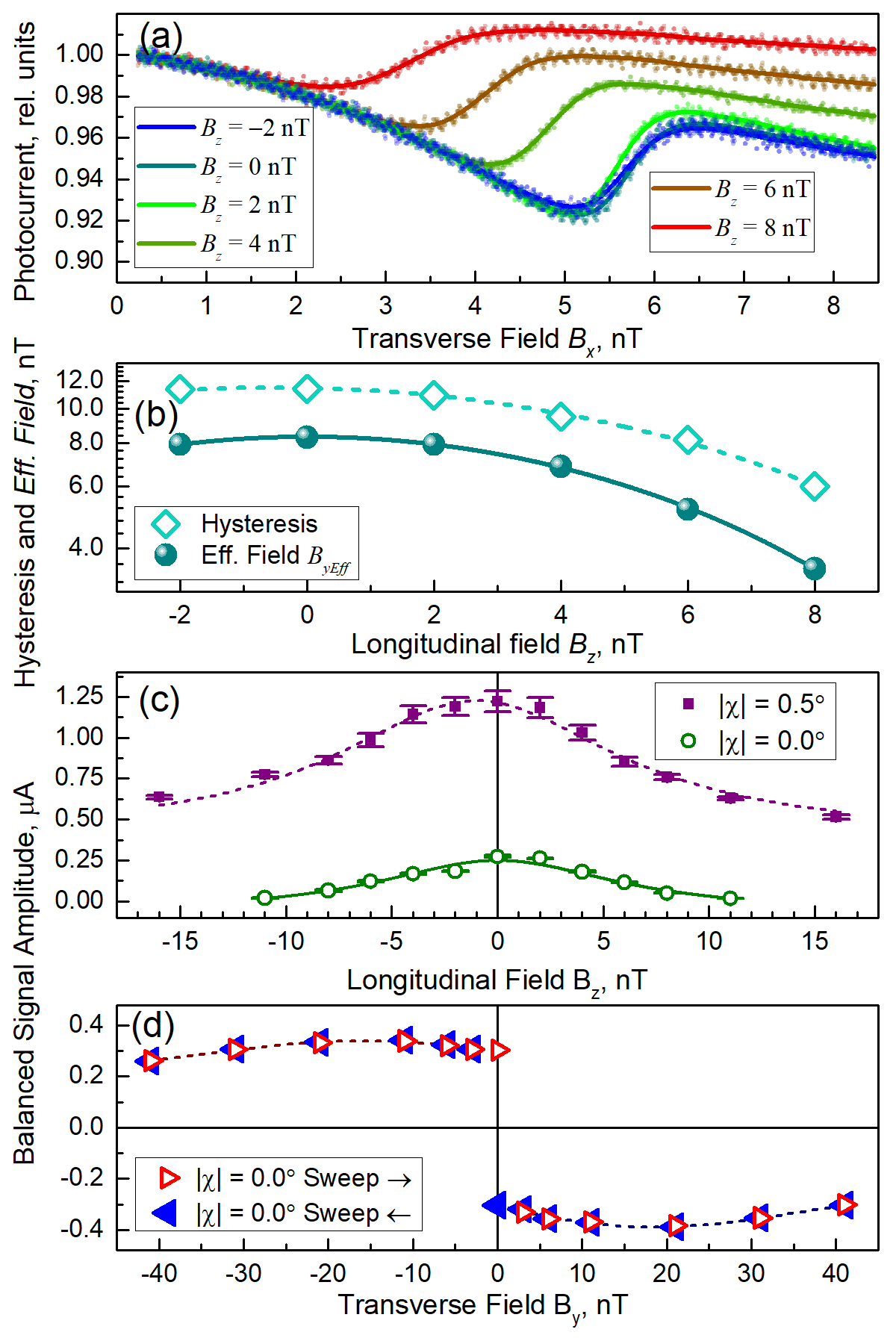}
	\caption{(a) Examples of recordings of transient processes (signal $S_T$) for different values of the longitudinal field $B_z$ and the ellipticity angle $\chi$ = +0.$5^\circ$; (b) the results of their processing -- hysteresis and the magnitude of the effective field, dependences of signal amplitudes and the corresponding relaxation times (inverse resonance widths) on the longitudinal field $B_z$; (c) amplitudes of the symmetric and antisymmetric resonance contours in Fig.~\ref{fig3}(c). Solid lines are approximations by Lorentz contours; (d) resonance amplitudes Fig.~\ref{fig3}(d). Solid lines are approximations by polynomials.}\label{fig5}
\end{figure} 
The results of processing the data presented in Fig.~\ref{fig3} are shown in Fig.~\ref{fig4} and Fig.~\ref{fig5}. Fig.~\ref{fig4}(a),(b) show the widths $\Gamma_{S}$, $\Gamma_{A}$ and amplitudes $A_{S}$, $A_{A}$ of both contours -- symmetric (the alignment signal) and antisymmetric (the orientation signal). The widths increase linearly with the power of the circular component $P_{\text{C\,in}}=P_{\text{in}}\sin{|2\chi|} \textcolor{Black}{\sim |\chi|}$. 

\textcolor{Black}{It can be seen that the graphs Fig.~\ref{fig4} are characterized by some shift along the horizontal axis; for the graphs in Fig.~\ref{fig4}(a), the weighted average value of this shift is $\chi_0$ = (-0.04 $\pm$ 0.01)$^\circ$. According to the specifications, the ellipticity accuracy of the Thorlabs polarimeter is $\pm0.25^\circ$, so this offset can easily be attributed to the error of this instrument. After taking this shift into account, we obtain the dependences of the widths on the ellipticity in the form} 

\textcolor{Black}{$\Gamma_{S} = (8.13 \pm 0.16)\text{~nT} + \left((5.26 \pm 0.25)\text{~nT}/^\circ \right)|\chi - \chi_0|$,}

\textcolor{Black}{$\Gamma_{A} = (10.9 \pm 0.1)\text{~nT} + \left((4.54 \pm 0.08)\text{~nT}/^\circ \right)|\chi - \chi_0|$,}

\noindent \textcolor{Black}{
 and the difference between the parameters of the left and right branches calculated separately lies within the margin of error. We see that when accounting for the zero offset, the width asymmetry completely disappears within the error limits. Since the decomposition of the signal into symmetric and antisymmetric contours with widths $\Gamma_{S}$ and $\Gamma_{A}$ is the basis of the proposed model, such symmetry, in our opinion, further confirms this model.}

\textcolor{Black}{This, however, is not entirely true for the hysteresis H. The asymmetry of the slope of the two branches of the inverse hysteresis is greater than the error (more details are given in Appendix~\hyperref[appendix:B]{B}). When processing the data in Fig.~\ref{fig3}, it was also found that the dependence of the widths $\Gamma_{S}$ and $\Gamma_{A}$ on the scanning direction of the $B_x$ field was absent within the error limits, but the position of the hysteresis loop could shift significantly depending on the ellipticity (see Fig.~\ref{fig3}(b)). This could be due to a number of factors, including the presence of the $B_y$ field ($2-3$~nT), the presence of a residual uncompensated $B_z$ field, and the influence of light shift. Unfortunately, complete suppression of these factors requires experimental techniques significantly superior to ours.}

 The factors contributing to the broadening of the resonances are two-fold: firstly, the broadening due to the pump circular light component, and secondly, the broadening due to the SERF mode destruction in the presence of a magnetic field $\bf{B}$ which arises due to the light shift at non-zero ellipticity. For $P_{\text{in}} \approx $ 2~mW and $|\chi| \le 1^{\circ}$, the total broadening is estimated as 4.0~nT/$^\circ$, which matches the data in Fig.~\ref{fig4}(a) closely \textcolor{Black}{(for details see Appendix~\hyperref[appendix:C]{C})}.

The amplitude of the antisymmetric contour is proportional to $P_{\text{C\,in}}$, and the amplitude of the symmetric contour is proportional to the intensity of the linear component, which in this experiment changes by no more than $2\%$.

\textcolor{Black}{\section{Discussion}\label{sec:5}}
\textcolor{Black}{Let us try to interpret the obtained data. Firstly, the analysis of the dependences on the $B_z$ and $B_y$ fields of the signals recorded during scanning of the $B_x$ field  (Fig.~\ref{fig3}(c),(d)), and the comparison of the symmetry of these signals with \eqref{eq:eq2} allow us to unambiguously state that, at zero pump ellipticity, we, as in \cite{Petrenko_Vershovskii_2025}, are dealing precisely with alignment signals. Furthermore, the widths of these signals ($8-10$ nT at $T_c = 145-150 ^\circ $C) clearly indicate the SERF mode. The introduction of ellipticity leads to the appearance of orientation signals, also in the SERF mode.  More than that, the orientation and alignment in the SERF mode coexist in the cell volume.} 

\textcolor{Black}{We will start by considering the possible causes of bistability, which is expressed in the inversion of a symmetrical demodulated $S_B$ contour.} Due to symmetry, the introduction of ellipticity cannot lead to re-pumping of the alignment, i.e. to a change in the sign of $\rho_{0}^{(2)}$. This is confirmed by a comparison of the signals $S_T \sim \rho_{0}^{(2)}$ and $S_B \sim \rho_{\pm1}^{(2)}$, where the former do not demonstrate inversion (Fig.~\ref{fig3}(a),(b)). Consequently, ellipticity in the system leads to a change in the sign of the alignment signal, caused by the multipole $\rho_{\pm1}^{(2)}$ \cite{Meraki_Elson_Ho_Akbar_Kozbial_Kolodynski_Jensen_2023} -- the coherence signal “frozen” in the ultraweak field. As previously mentioned, this sign is determined by the field component $B_y$. Ellipticity introduces two significant factors into the system: a virtual longitudinal magnetic field (i.e. light shift), and orientation. Light shift can be easily simulated in the experiment by introducing a $B_z$ field into the system. As follows from Fig.~\ref{fig3}(c), at $\chi = 0^\circ$ the field $B_z$ is unable to invert the alignment signal; instead, it changes the alignment signal’s shape, as predicted by the theory. Therefore, the inversion is due to orientation.

Indeed, elliptical pumping creates a moment $\bf{M}^{\text{(1)}}$ (the index "1" corresponds to orientation, "2" to alignment) along the \emph{z} axis. In the presence of a nonzero field $B_x$ (i.e., during scanning), this moment rotates around the \emph{x} axis, resulting in a component of the moment $M^{(1)}_y$ (Fig.~\ref{fig2}(d)). The sign of  $M^{(1)}_y$ is determined by the sign of the ellipticity and the sign of $B_x$. If we assume that the moment $\bf{M^{(1)}}$ is capable of creating an effective field  $\bf{B_{\text{Eff}}}$ parallel to it, then it is capable of changing the sign of the alignment signal when $B_{y\,\text{Eff}}+B_{y}$ passes through zero. We note once again that $B_z$, and therefore $M^{(1)}_z$, are not capable of inverting the alignment signal.

The question arises as to how the effective field $\bf{B_{\text{Eff}}}$ can arise. It is known that in a symmetric cell, the integral field produced by identically oriented, uniformly distributed moments within the cell vanishes (which is why atomic magnetometers are not subject to shifts due to the magnetization of the atomic ensemble). 

But the condition of uniform distribution is violated in an optically dense medium, that is, precisely under those conditions -- on the verge of complete absorption of the pumping -- under which we observe the bistability effect. If the center of the ensemble of oriented atoms is shifted relative to the center of the ensemble of aligned atoms (which is due to a significantly stronger absorption of the linear component of the pumping), then the latter will be affected by the magnetic field created by the former. In this case, it is possible that the \textcolor{Black}{stretched} ensembles of alignment and orientation are spatially separated. Then, due to the SERF effect, they will be concentrated in local, clearly defined areas, and this will allow them to coexist. 

A simple estimate shows that to create a field of ${B_{\text{at}} = 1~\text{nT}}$ at a distance of $d = $ 1~mm, a point dipole containing ${N_{\text{at}} \approx 5\times10^{11}}$ oriented atoms is required. Under our conditions, $N_{\text{at}}$ will occupy a cube with a side of $L \approx $ 1.3 mm, so $d/L\approx$~0.8 \textcolor{Black}{(note that the ratio $d/L$ does not depend on $d$)}.
The condition for treating the dipole as point-like can be written as $d/L \gg 2$. It is evident that it is not fully satisfied, but one can see that the fields created by the atoms are close in order of magnitude to the required ones. This can explain why the effect is observed in such a narrow temperature range: at lower temperatures, the concentrations of atoms are insufficient to create the necessary field, while at higher temperatures, complete absorption of light in the cell occurs. 

\textcolor{Black}{It should be emphasized that the authors do not insist on the proposed model of spatial separation of the alignment and orientation areas. This model is preliminary and requires further experimental and theoretical validation; it was put forward mainly to demonstrate the fundamental possibility of the coexistence of orientation and alignment in one "stretched" atomic ensemble and the interaction between them. Alternatively, the effective field may arise as a result of Fermi contact interaction \cite{Grover_1978}. However, for the Cs-Cs pair such an interaction is highly unlikely: it should be hampered by the repulsion of electron shells and spin-exchange interaction. It also remains possible that, in zero field, collective effects (namely, extremely fast spin-exchange and slow diffusion under high buffer gas pressure) result in a local, submillimeter-scale spontaneous transformation of alignment into orientation, analogous to the formation of domains in a ferromagnet below the Curie point \cite{Klipstein_Lamoreaux_Fortson_1996, Petrenko_Vershovskii_2025}. The observed alignment effects are then due to the dynamics of coexisting domains with opposite orientations, and the coexistence of alignment and orientation is explained by a change in the quantitative ratio and size of such domains. This issue requires further study.} 

The presence of hysteresis (Fig.~\ref{fig4}(c)) is a manifestation of collective SERF effects. \textcolor{Black}{Here, it's worth recalling the pioneering studies \cite{Fortson_Heckel_1987, Klipstein_Lamoreaux_Fortson_1996}, which investigated the longitudinal polarization of Cs in nonzero magnetic fields. A direct analogy can be drawn: since longitudinal polarization is unaffected by Larmor precession, its dynamics are similar to the dynamics of the complete polarization of an ensemble in a zero magnetic field. The collective effects in these two cases may also be similar. Indeed, Fortson and colleagues showed that the probability of an atom occupying a state with the maximum angular momentum modulus is greater the larger the fraction of atoms already in this state. When pumped with linearly polarized or unpolarized light, this can lead to spontaneous polarization of the atomic medium. When pumped with elliptically polarized light, hysteresis loops were observed in the \textit{orientation} signal, vaguely similar to those we observed in the \textit{alignment} signals (although in our case, to invert the alignment signals, it is sufficient to change the ellipticity by tenths of a degree, rather than tens of degrees, or the magnetic field by $1-2$~nT).}

However, the numerous experiments investigating pure orientation in the SERF mode during field scanning \cite{Li_Quan_Zhou_Wang_Lu_Hu_Liu_Fang_2018, Savukov_2010, Savukov_2017} have failed to note any unusual hysteresis effects. Therefore, it must be assumed that either the inversion hysteresis is inherent to alignment, or alignment is capable of hindering the oriented moments rotation in the magnetic field $B_x$ up to a certain point.

It can be hypothesized that the inversion occurs when the component $M^{(1)}_y$ reaches a certain threshold value $\pm M^{(1)}_{y0}$: then, in the first approximation,
\begin{equation}
    M^{(1)}_{y0} = M^{(1)} {\text{sin}}\phi \approx M^{(1)} \gamma B_{x0} /\Gamma,
    \label{eq:eq3}
\end{equation}
where $\phi$ is the angle of rotation of the moment $\bf{M}^{(1)}$, $\Gamma$ is the resonance width, $\pm B_{x0}$ are the values of $B_x$ at which the transient process begins. Then $B_{x0} \sim \Gamma(\chi)/\chi$; and indeed, in our experiment, the hysteresis dependence on $\chi$ is close to hyperbolic (Fig.~\ref{fig4}(c)). The slight difference in the dependence of the hysteresis from the hyperbola can be explained by the dependence $\Gamma(\chi)$.

Figures~\ref{fig4} (d),(e) show the transient process velocity and the presumed effective field $B_{y\,\text{Eff}}$, which could lead to the reversal of the alignment moment by $\pi$ during the time corresponding to the process duration $\Delta t$: ${B_{y\,\text{Eff}} = 1/(\Delta t(\gamma/2\pi))}$. We see that $|B_{y\,\text{Eff}}|$ is almost independent of the ellipticity, and in our case is equal to $\sim $1.1~nT (in the estimation, the pumping was assumed to be weak, which in zero fields corresponds to $\gamma/2\pi =$ 1.27~Hz/nT; under strong pumping $\gamma/2\pi =$ 3.5 Hz/nT). This confirms the hypothesis that the inversion occurs at the moment when the field $B_{y\,\text{Eff}}$ reaches the threshold value. The characteristic values of the ellipticity angle, at which the value of $B_{y\,\text{Eff}}$ approaches saturation, are $\chi \approx 0.1^\circ$. The amplitude of the orientation signals becomes equal to the amplitude of the alignment signals at the ellipticity angle value of $\chi \approx 0.2^\circ$ (Fig.~\ref{fig4}(b)).

Fig.~\ref{fig5}(b) shows the dependences of the hysteresis and effective field on $B_z$. Both of them are described by Lorentz contours with close widths ($25-32$~nT). Analysis of the signal amplitudes in Fig.~\ref{fig5}(c),(d) and their comparison with the theory \cite{Meraki_Elson_Ho_Akbar_Kozbial_Kolodynski_Jensen_2023} shows that their dependence on the $B_y$ field is significantly smoothed, as if the $B_y$ field is stabilized by external factors. At the same time, at $B_y = 0$, a sharp inversion of the signals is observed, also accompanied by bistability. From this, we can conclude that the magnitude of the effective total field along the \emph{y} axis in the system is maintained at the same level, but with different signs, due to collective effects. This is an unexpected effect that requires study.

\textcolor{Black}{\section{Conclusions}\label{sec:6}}
Thus, we have demonstrated that alignment and orientation can coexist in the SERF regime, while maintaining a distinction in their dynamics. This result is nontrivial, since the SERF regime itself assumes the dominance of a single momentum state, and the possibility of SERF in alignment has not yet been theoretically substantiated. We have also shown that the dipole and quadrupole moments, which, according to classical theory, should evolve independently, interact; this interaction is nonlinear and leads to the emergence of bistability in the medium. We have proposed a qualitative explanation of the mechanism of such interaction and shown that the alignment inversion is determined by the transverse component of orientation. However, the underlying details of the process remain unresolved, and we hope that they will attract the attention of theoreticians. In particular, the role of buffer nitrogen, which is characterized by a non-zero orbital molecular momentum \cite{Andalkar_Warrington_Romalis_Lamoreaux_Heckel_Fortson_2002}, is not fully understood. Nevertheless, these results have clear practical significance, since they demonstrate the fundamental possibility of creating optical switches and memory cells on atomic vapors. Such devices cannot be characterized by high speed, but are capable of demonstrating a record storage time for thermal atoms -- hundreds of seconds or more. In the future, they can be used to demonstrate a long-lived quantum entangled state of two ensembles. Essentially, these elements can be two-input: they can be controlled by changing both the magnetic field and the ellipticity of the input light (a change of $0.1-0.2^\circ$ is sufficient), and their output signal is a rotation of the plane of polarization of the output light (by $1-2^\circ$). This rotation can easily be converted into a change in ellipticity for use after optical amplification in the next element. The effect can also be used to detect small ($\le 0.1^\circ$) values of the ellipticity of the resonant light.

\textbf{}

\textbf{Conflict of interest:} The authors declare that they have no conflict of interest.

\textbf{}

\textbf{Funding:} This research was funded by the baseline project FFUG-2024-0039 at the Ioffe Institute.

\textbf{}

\textbf{Acknowledgments:} The authors thank Prof. Eugene Aleksandrov and Dr. Anatoly Pazgalev for valuable discussions.

\appendix
\textcolor{Black}{
\appendix\section*{Appendix A}\label{appendix:A}}
\addcontentsline{toc}{section}{Appendix}

\noindent\textcolor{Black}{\textbf{Light absorption and signals in the cell.}
This section contains additional details regarding the experimental conditions. At 2~mW of input light power the typical photocurrent value with a photodetector efficiency of $\sim$0.7~A/W was $I_{\text{ph}} \approx$ 6~$\mu$A with linear polarization and $I_{\text{ph}} \approx$ 140~$\mu$A with circular polarization of pump light. This means that linearly polarized light was typically attenuated from 2~mW to 8.5~$\mu$W, or 230 times. Circularly polarized light was attenuated to 200~$\mu$W, corresponding to a tenfold reduction.}

\textcolor{Black}{The typical alignment signal value was $S_{\text{Al}} \approx$ 0.3~$\mu$A, which corresponded to a polarization rotation of $\sim2^\circ$. For circular polarization, the corresponding values are $I_{\text{ph}} \approx$ 140~$\mu$A and $S_{\text{Or}} \approx$ 100~$\mu$A.}

\textbf{}

\textcolor{Black}{
\appendix\section*{Appendix B}\label{appendix:B}}
\addcontentsline{toc}{section}{Appendix}

\noindent\textcolor{Black}{\textbf{The results of approximation.}
Below are the results of separate approximation of the left and right parts of of the widths $\Gamma_{S}$ and $\Gamma_{A}$ (Figure~\ref{fig4}(a)) and the inverse hysteresis $H^{-1}$ (Figure~\ref{fig4}(c)):}

\textcolor{Black}{$\Gamma_{SL} = (8.07 \pm 0.18)\text{~nT} + \left((5.25 \pm 0.31)\text{~nT}/^\circ \right)|\chi - \chi_0|$,}

\textcolor{Black}{$\Gamma_{SR} = (8.19 \pm 0.10)\text{~nT} + \left((5.27 \pm 0.17)\text{~nT}/^\circ \right)|\chi - \chi_0|$,}

\textcolor{Black}{$\Gamma_{AL} = (10.92 \pm 0.08)\text{~nT} + \left((4.55 \pm 0.13)\text{~nT}/^\circ \right)|\chi - \chi_0|$,}

\textcolor{Black}{$\Gamma_{AR} = (10.89 \pm 0.02)\text{~nT} + \left((4.54 \pm 0.04)\text{~nT}/^\circ \right)|\chi - \chi_0|$,}

\textcolor{Black}{$H_{L}^{-1} = (0.10 \pm 0.02)\text{~nT} + \left( (0.32 \pm 0.02)\text{~nT}/^\circ \right)|\chi - \chi_0|$,}

\textcolor{Black}{$H_{R}^{-1} = (0.08 \pm 0.02)\text{~nT} + \left( (0.44 \pm 0.03)\text{~nT}/^\circ \right)|\chi - \chi_0|$,}

\noindent\textcolor{Black}{
 where $L$ and $R$ indices correspond to the left and right branches, and $\chi_0$ = (–0.04 $\pm$ 0.01)$^\circ$.}

\textbf{}

\textcolor{Black}{
\appendix\section*{Appendix C}\label{appendix:C}}
\addcontentsline{toc}{section}{Appendix}
\noindent\textcolor{Black}{\textbf{The broadening of the resonance lines.}}
 The expected broadening of the resonances was calculated as follows: the broadening due to the pump circular light component is equal to $P_{\text{C\,in}} k_{\text{LB}}$, where for our conditions $k_{\text{LB}} = (30~\pm~3)$~nT/mW (measured at $|\chi| \approx$ 4$5^\circ$). 
 
 The broadening due to the SERF mode destruction in the presence of a magnetic field $\bf{B}$ in the first order is equal to $k_{\text{SERF}} |\bf{B}|$, \textcolor{Black}{where  $k_{\text{SERF}}$ is the linear term of the dependence of the magnetic resonance width on the magnetic field in the SERF mode} (for our conditions $k_{\text{SERF}} \approx$ 0.3~nT/nT \cite{Happer_Tam_1977, Petrenko_Vershovskii_2025}). 
 
 When ellipticity appears, the field $\bf{B}$ arises due to the light shift. The latter is equal to $k_{\text{LS}} P_{\text{C\,in}}$, where  $k_{\text{LS}} = $ (90~$\pm$~5)~nT/mW (measured at $|\chi| \approx$ 4$5^\circ$). 
 
 For $P_{\text{in}} \approx $ 2~mW and $|\chi| \le 1^{\circ}$, the total broadening is estimated as 4.0~nT/$^\circ$.

%\begin{thebibliography}{}

%\end{thebibliography}

\bibliography{bibl}
\end{document}